\documentclass[fleqn,usenatbib]{mnras}

\usepackage{newtxtext,newtxmath}
\usepackage[T1]{fontenc}

\DeclareRobustCommand{\VAN}[3]{#2}
\let\VANthebibliography\thebibliography
\def\thebibliography{\DeclareRobustCommand{\VAN}[3]{##3}\VANthebibliography}

\usepackage{graphicx}	% Including figure files
\usepackage{amsmath}	% Advanced maths commands
\title[Dawn storms and X-rays]{Jupiter's X-ray aurora during UV dawn storms and injections as observed by XMM-Newton, Hubble, and Hisaki}

\author[Affelia D. Wibisono et al.]{
A. D. Wibisono$^{1,2}$\thanks{E-mail: affelia.wibisono.18@ucl.ac.uk},
G. Branduardi-Raymont$^{1,2}$,
W. R. Dunn$^{1,2}$,
T. Kimura$^{3*}$,
A. J. Coates$^{1,2}$,
D. Grodent$^{4}$,
\newauthor
Z. H. Yao$^{5}$,
H. Kita$^{6}$,
P. Rodriguez$^{7}$,
G. R. Gladstone$^{8}$,
B. Bonfond$^{4}$
and R. P. Haythornthwaite$^{1,2}$
\\
$^{1}${Mullard Space Science Laboratory, Department of Space \& Climate Physics, University College London, Holmbury St. Mary, Dorking, Surrey, RH5 6NT, UK}\\
$^{2}${The Centre for Planetary Science at UCL/Birkbeck, Gower Street, London, WC1E 6BT, UK}\\
$^{3}${Frontier Research Institute for Interdisciplinary Sciences Tohoku University, Japan}\\
$^{4}${Laboratoire de Physique Atmospherique et Planetaire, Universite de Liege, Liege, B-4000, Belgium}\\
$^{5}${Key Laboratory of Earth and Planetary Physics, Institute of Geology and Geophysics, Chinese Academy of Sciences, Beijing, China}\\
$^{6}${Department of Information and Communication Engineering, Tohoku Institute of Technology, Japan}\\
$^{7}${European Space Astronomy Centre, Madrid, Spain}\\
$^{8}${Southwest Research Institute, San Antonio, Texas, USA}\\
$^{*}${Now at Department of Physics, Faculty of Science, Tokyo University of Science, Japan}\\
}

\date{Accepted XXX. Received YYY; in original form ZZZ}

\pubyear{2021}

\begin{document}
\label{firstpage}
\pagerange{\pageref{firstpage}--\pageref{lastpage}}
\maketitle

\begin{abstract}
We present results from a multiwavelength observation of Jupiter’s northern aurorae, carried out simultaneously by XMM-Newton, the Hubble Space Telescope (HST), and the Hisaki satellite in September 2019. HST images captured dawn storms and injection events in the far ultraviolet aurora several times during the observation period. Magnetic reconnection occurring in the middle magnetosphere caused by internal drivers is thought to start the production of those features. The field lines then dipolarize which injects hot magnetospheric plasma from the reconnection site to enter the inner magnetosphere. Hisaki observed an impulsive brightening in the dawnside Io plasma torus (IPT) during the final appearance of the dawn storms and injection events which is evidence that a large-scale plasma injection penetrated the central IPT between 6-9 R\textsubscript{J} (Jupiter radii). The extreme ultraviolet aurora brightened and XMM-Newton detected an increase in the hard X-ray aurora count rate, suggesting an increase in electron precipitation. The dawn storms and injections did not change the brightness of the soft X-ray aurora and they did not “switch-on” its commonly observed quasi-periodic pulsations. Spectral analysis of the X-ray aurora suggests that the precipitating ions responsible for the soft X-ray aurora were iogenic and that a powerlaw continuum was needed to fit the hard X-ray part of the spectra. The spectra coincident with the dawn storms and injections required two powerlaw continua to get good fits.
\end{abstract}


%%%%%%%%%%%%%%%%%%%%%%%%%%%%%%%%%%%%%%%%%%%%%%%%%%

%%%%%%%%%%%%%%%%% BODY OF PAPER %%%%%%%%%%%%%%%%%%

\section{Introduction}

The first extraterrestrial auroral emissions were detected in 1979 by Voyager 1 which led to the discovery of Jupiter's ultraviolet (UV) aurora (\cite{Broadfoot}). In the same year, the gas giant's X-ray aurora was identified by using data from the Einstein Observatory (\cite{Metzger}). The brightest region of Jupiter's aurora is the main auroral oval, also known as the main emission. This oval is permanent and is produced by a population of energetic electrons. Some of these precipitating electrons excite molecular and atomic hydrogen in Jupiter's atmosphere that will subsequently release UV photons when they return back to the ground state. Other electrons emit high energy (“hard”) X-rays via bremsstrahlung radiation when they are slowed and deflected by native molecules (\cite{GBRStudy}). Diffuse UV and low energy (“soft”) X-rays can be found poleward of the main oval (\cite{GBRSpec}). Charge stripping produces high charged state ions that then charge exchange with neutrals in the planet's atmosphere to produce the soft X-ray emissions (e.g. \cite{Cravens,GBRFirst,Bhardwaj2005,Elsner,Gladstone1998,Kimura,Dunn2020a}). There seems to be a preference to an iogenic source for these ions according to observational and theoretical studies (e.g. \cite{Cravens,Hui2009,Dunn2016}), however, the inclusion of solar wind ions is sometimes needed to fit the auroral spectrum (\cite{Dunn2020b,Hui2010}). 

Jupiter's aurorae respond to changing conditions within and outside of Jupiter's huge magnetosphere (\cite{Grodent}). For example, the dusk sector of the main emission dims and thins in the UV waveband when the magnetosphere is expanded and contains very little plasma. The UV aurora has been observed to brighten with solar wind compression events (\cite{Clarke,Nichols2009,Nichols2017}) and the increase in the total power of the aurora has a stronger correlation with the time between the compressions rather than the amplitude in the solar wind dynamic pressure increase (\cite{Kita2016,Kimura2019}). During solar wind compressions, the main UV oval also becomes bright and well defined in the dawn sector. Features in the UV aurora that are caused by solar wind compressions and rarefactions tend to persist and/or evolve for several Jupiter rotations. \cite{Dunn2020b} found that the X-ray aurora can brighten during solar wind compressions and during intervals of quiet solar wind which suggests that the aurora can be controlled by processes happening inside the magnetosphere, or by the direction of the interplanetary magnetic field that would allow for dayside reconnection to happen.

\subsection{Dawn Storms and Injection Events}
Earth-orbiting observatories, such as the Hubble Space Telescope (HST) can only witness features of Jupiter's aurora, such as dawn storms and injection events, between the dawn and dusk sectors. However, during perijove, Juno's Ultraviolet Spectograph (UVS) instrument can provide images of this phenomena from every local time sector. \cite{Bonfond2020} used Juno's first 20 orbits to show that dawn storms tend to start as transient spots in the pre-midnight sector. A few hours later, the main emission at midnight brightens and beads may form. This bright arc in the main emission continues to brighten and thicken as it moves towards the dawn sector to become a dawn storm. The arc may split into two branches with one moving polewards. Both branches will dim after a few hours and the part of the dawn storm that is at lower latitudes will evolve into a distinct patchy enhancement between the main oval and Io's footprint towards the dusk sector: this is what is called an injection event. The whole process takes 5-10 hours to complete, however, chains of multiple dawn storms have also been observed, as well as dawn storms without accompanying injection events and vice versa. \cite{Bonfond2020} also show that Jupiter's dawn storms share many of the same signatures as the Earth's auroral substorms. Dawn storms are thought to be related to internally driven explosive reconfigurations of the magnetotail and generally starting with reconnection events in the distant magnetosphere ($\sim$90 R\textsubscript{J} (Jupiter radii)). After reconnection, the magnetic field lines undergo dipolarization which causes strong injections of hot magnetospheric plasma from the middle magnetosphere into the inner magnetosphere to produce the injection events (\cite{Yao2020,Haggerty2019,Dumont2018,Mauk2002}). Furthermore, the formation of dawn storms seems to be independent of solar wind compressions (\cite{Yao2020,Bonfond2020,Kimura2015,Nichols2009}). 

HST witnessed a dawn storm in March 2017. At this time, Juno was $\sim$80-60 R\textsubscript{J} away from the planet and was in the equatorial plane (\cite{Yao}). The spacecraft’s JEDI, Waves and Magnetometer Investigation (MAG) instruments showed two instances of magnetic loading and unloading at this distance which correlated well with electron energization and cooling. Furthermore, the HST images showed bright auroral emissions at the start of the unloading processes, whereas the aurora was relatively dim during the loading processes. \cite{Yao} also showed that magnetic reconnection occurred during the magnetic loading and unloading. Although these magnetic processes occurred at $\sim$60-80 R\textsubscript{J} they may still affect auroral enhancements in the main oval which map to field lines located at 20-30 R\textsubscript{J}. \cite{Yao} made two suggestions to explain this. The first is that the majority of auroral precipitation originates at 20-30 R\textsubscript{J}, but there may also be comparable trends at 60-80R\textsubscript{J}. Their second suggestion involves a current loop system between 20-30 R\textsubscript{J} and 60-80 R\textsubscript{J}. The unloading processes detected by Juno at 60-80 R\textsubscript{J} could enhance downward currents formed at this distance. The corresponding upward currents at 20-30 R\textsubscript{J} should then also enhance in response and cause the aurora to brighten. Another set of simultaneous observations by HST and Juno in May 2017 suggested that dawn storms and injection events are physically connected to each other (\cite{Yao2020}).

Signatures in the X-ray aurora due to reconnection, mass loading and injection events are currently unknown.

\subsection{Instrumentation}
\begin{figure}
    \includegraphics[width=\columnwidth]{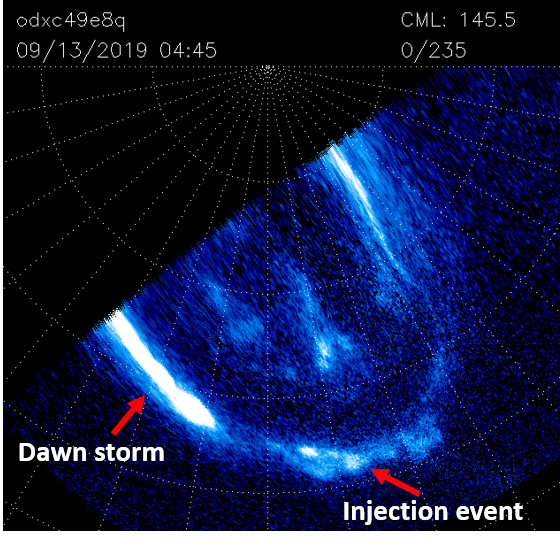}
    \caption{Hubble Space Telescope (HST) polar projection image of the far UV (FUV) northern aurora taken in September 2019 with a dawn storm and injection event labelled.}
    \label{fig:DawnStorm}
\end{figure}

Observations of Jupiter's far UV (FUV) aurora by HST were undertaken by using its Space Telescope Imaging Spectrograph (STIS) instrument. It has a spatial resolution of 0.08 arcsec (\cite{Grodent2003}) and detects FUV auroral emissions with energies 7.29-9.92 eV (wavelengths of 1250-1700 Å). Dawn storms appear in Hubble Space Telescope (HST) images as brightenings in the FUV dawnside main oval emission (see Fig.~\ref{fig:DawnStorm}). A series of observations taken over several HST orbits around the Earth show that these features appear and disappear over one Jupiter rotation.

Hisaki's payload consists of the Extreme Ultraviolet Spectroscope for Exospheric Dynamics (EXCEED) instrument (\cite{Yoshioka,Yoshikawa,Yamazaki}) which produces spectral images in the energy range of 8.4-23.8 eV (1480-520 Å). The spectral resolution in this energy range is 3.0-5.0 Å at Full Width at Half Maximum (FWHM) and a spatial resolution of 17 arcsec can be achieved. EXCEED has a time resolution of 1 minute and can observe for up to 50 minutes for every one of its 106 minute-long orbits around the Earth. During our observations, the centers of two dumbbell slits with widths of 20 and 140 arcsec and lengths of 360  arcsec were positioned over Jupiter's northern aurora while the ansae captured the dawn and dusk sides of the Io Plasma Torus (IPT).

XMM-Newton is an Earth-orbiting multiwavelength observatory that has a payload of three X-ray telescopes and the Optical Monitor (\cite{Mason}) which detects optical and UV wavelengths. Jupiter is a bright optical object, therefore, the Optical Monitor must be closed during Jupiter observations to prevent damaging the instrument. The European Photon Imaging Camera (EPIC) X-ray instrument consists of one pn CCD camera (\cite{Struder}) and two metal-oxide semiconductor (MOS) cameras (\cite{Turner}) that are sensitive between 0.15-12 keV (82.66-1.03 Å), and have spatial and spectral resolutions of 15 arcsec (at the half energy width) and 80 eV at 1 keV (0.99 Å at 12.40 Å), respectively. The EPIC instrument has a high sensitivity which is essential to capture the limited number of X-ray photons from Jupiter's aurorae. Using data from this instrument ensures that we receive sufficient counts for time-series analysis and so that we can compare the aurora from rotation-to-rotation.

\section{September 2019 Jupiter Observations}
\begin{figure}
    \includegraphics[width=\columnwidth]{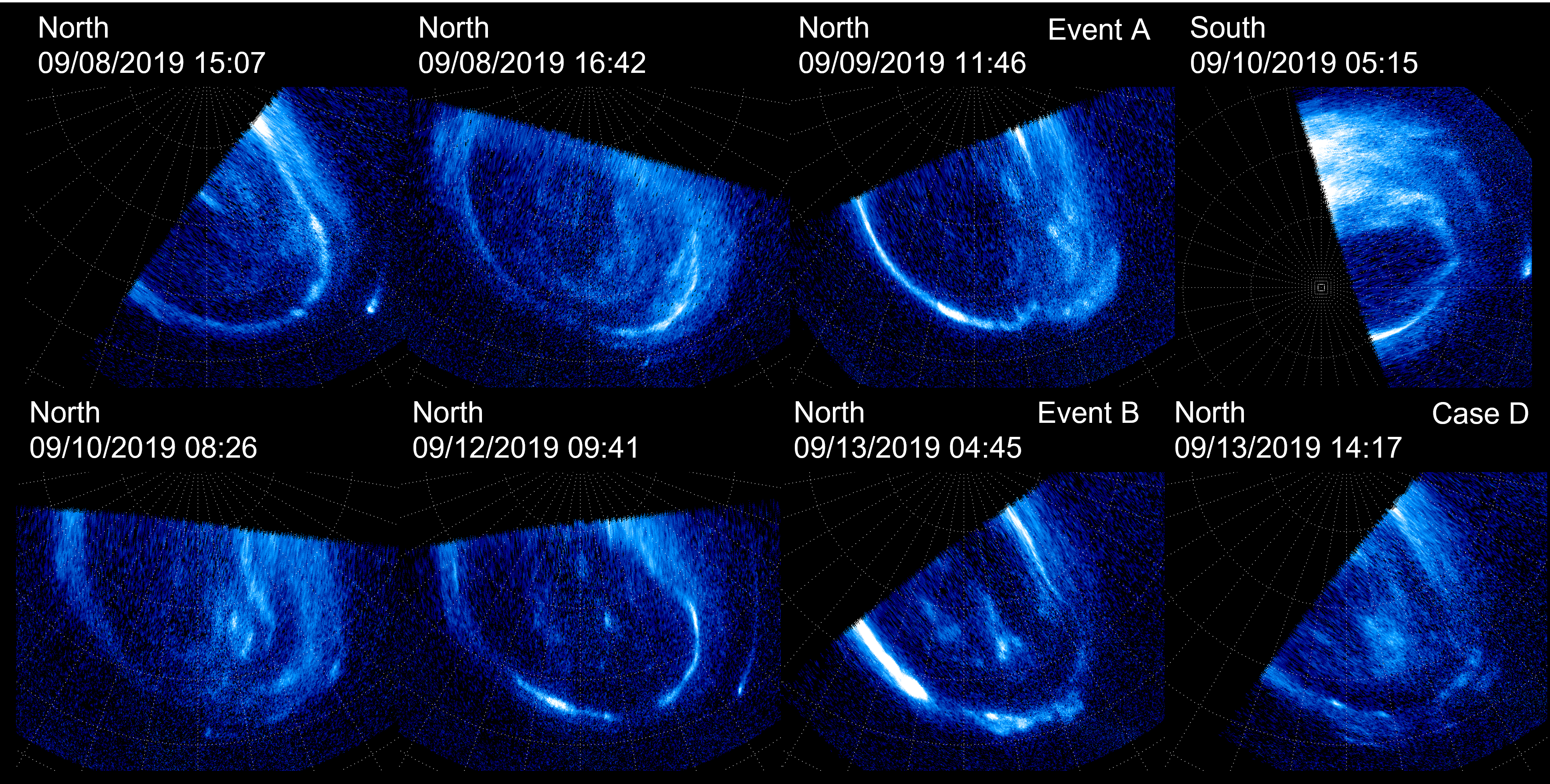}
    \caption{The HST images of Jupiter's northern and southern FUV aurora from 8-13 Sept 2019. Events A and B show the appearances of dawn storms and injection events in the northern aurora and Case D has neither of these features.}
    \label{fig:HSTImages}
\end{figure}

HST took 8 $\sim$40-minute-long observations of Jupiter's northern and southern FUV aurora between 8-13 Sept 2019 as part of the GO-15638 program (Fig.~\ref{fig:HSTImages}) that coincided with XMM-Newton's observations. Using the definitions of the morphological families in \cite{Grodent}, we conclude that most of the images show that the FUV aurora was in the quiet or unsettled groups. This suggests that the magnetosphere was largely undisturbed and contained very little plasma. However, faint emissions equatorward of the main oval (e.g. on 8 Sept 16:42 UTC) could be due to dipolarization of magnetic field lines beyond Europa's orbit that subsequently produces whistler mode waves. Electrons in the loss cone are scattered by these waves which then produce this auroral feature (\cite{Radioti2009}). Unfortunately, JEDI was not able to resolve the loss cone during Juno Perijoves 7-10 (July-December 2017) when auroral injections were detected (\cite{Haggerty2019}). However, JADE, JEDI and Waves data from Juno's first perijove show direct evidence for efficient pitch angle scattering to make electrons with energies 0.1-700 keV precipitate and produce diffuse auroral emissions located equatorward of the main oval and near the dusk sector. However, Juno did not detect whistler mode waves at this time, possibly because the spacecraft was in the polar region and too far from the magnetic equator which is where whistler mode waves tend to be detected at distances between 8-15 R\textsubscript{J} (\cite{Li}). We also see three occasions when the northern FUV aurora shows the presence of dawn storms and injection events. The first occurred on 9 Sept 11:46 UTC (Event A) which had a particularly large injection event in the dusk sector while the dawn storm is quite small and may be considered as a pseudo-dawn storm. The northern FUV aurora appeared quiet for the next few days until 12 Sept 09:41 UTC when another pseudo-dawn storm formed. However, XMM-Newton was at perigee at this time and was not gathering data. This pseudo-dawn storm then developed until a fully fledged dawn storm with an injection event half a Jupiter rotation later on 13 Sept at 04:45 UTC (Event B).

\begin{figure}
    \includegraphics[width=\columnwidth]{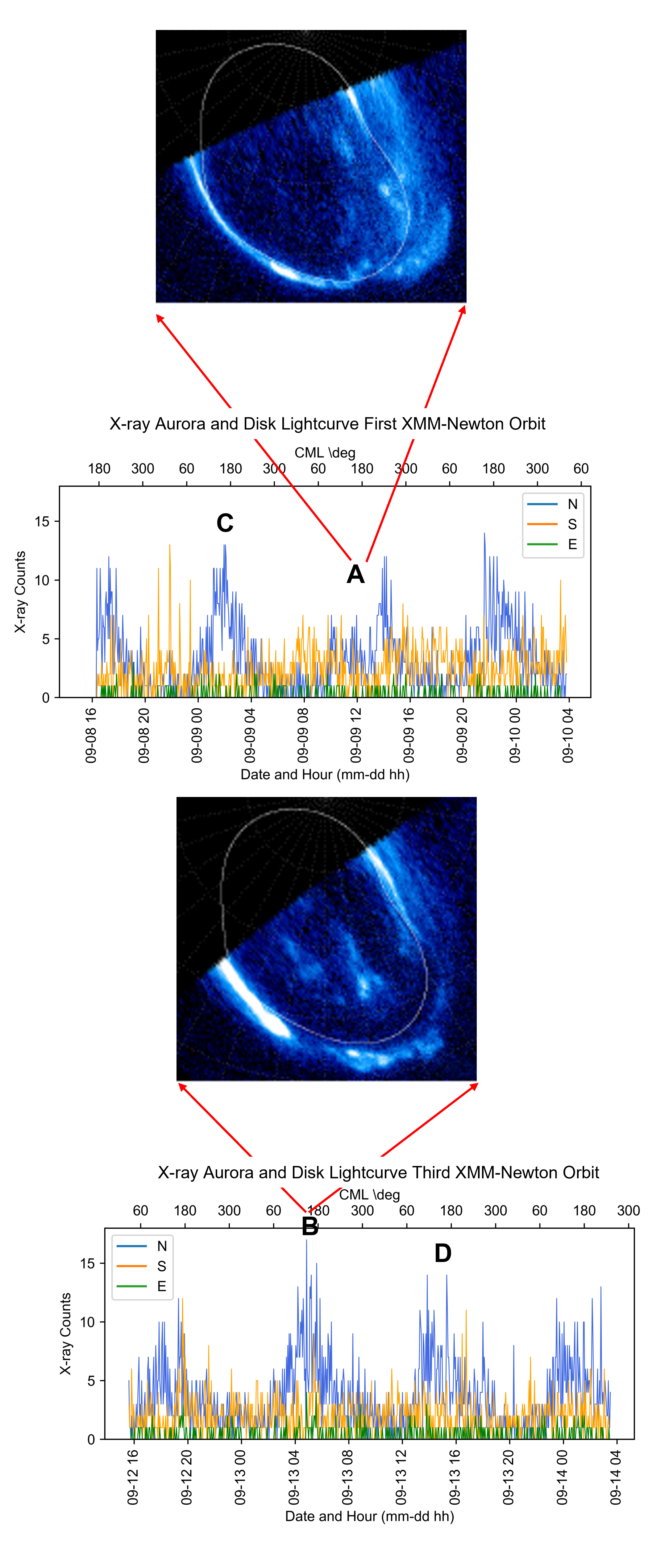}
    \caption{Three-minute binned XMM-Newton lightcurves of the northern and southern X-ray aurora are shown in blue and orange respectively while the disc emission is in green. They are produced by coadding data from the EPIC-pn and EPIC-MOS observations. The times of Events A and B are marked on the lightcurve and the HST images of those events are shown above the lightcurves. Cases C and D are also marked (see Section \ref{XRaySpectra}).}
    \label{fig:XMMLightcurves}
\end{figure}

XMM-Newton observed Jupiter for a total of $\sim$405 ks (6750 minutes) over 6 days which were split into 3 sets of observations with perigees in between as the spacecraft made 3 orbits around the Earth. This study will focus on XMM-Newton's first and third orbits around the Earth as this is when Events A and B appeared. The results from its second orbit will be presented in Wibisono et al., in prep. The XMM-Newton lightcurves of the northern (in blue) and southern (in orange) X-ray aurorae are shown in Fig.~\ref{fig:XMMLightcurves}. Also shown in green is the lightcurve of Jupiter's disc which is from solar X-rays reflected from the planet's equatorial regions. This emission has little variation and is always well below the levels of the auroral emissions meaning that the Sun did not release large X-ray flares. Based on this low emission, we do not expect the disc emission to contaminate the aurora significantly, therefore, we did not subtract it from the auroral data. The three different regions are defined on the EPIC-pn images of Jupiter in Fig.~\ref{fig:A1} of the Appendix.

Jupiter's $\sim$10 hour-long rotation is evident in the northern aurora lightcurve as the aurora rotates in and out of view due to it being fixed on Jupiter's frame (when the Central Meridian Longitude (CML) is 155-190\textdegree). The northern aurora's average brightness is fairly constant throughout the two orbits, but it seems to dim unexpectedly between $\sim$08:00-12:00 UTC on 9 September. The southern aurora is fixed between CMLs of 0-75\textdegree and its visibility is not in phase with that of the northern aurora. Its lightcurves from our observations are featureless apart from the relatively bright pulsating flares between $\sim$20:00-23:00 UTC on 8 September. The aurora in this hemisphere for the remainder of the observation was very dim with some short-lived and isolated flares which make it more difficult to identify the modulation from Jupiter's rotation.

The Hisaki Telescope was scheduled to observe Jupiter's aurora and the IPT throughout August and September so that would have allowed us to monitor any changes inside of Jupiter's magnetosphere. Unfortunately, Hisaki's star tracker sensor has suffered degradation which has caused it to have attitudinal problems from mid-2016 to the present day. As a consequence of this, Hisaki was not able to keep Jupiter's location in the slits stable throughout the observation period and the auroral and IPT powers were not recorded between 7-11 September. The extreme UV (EUV) auroral and IPT powers needed to be analysed with care because changes in Jupiter's location in the slit can result in artificial dimming or brightening in those values.

\section{Results}

\begin{figure}
    \includegraphics[width=\columnwidth]{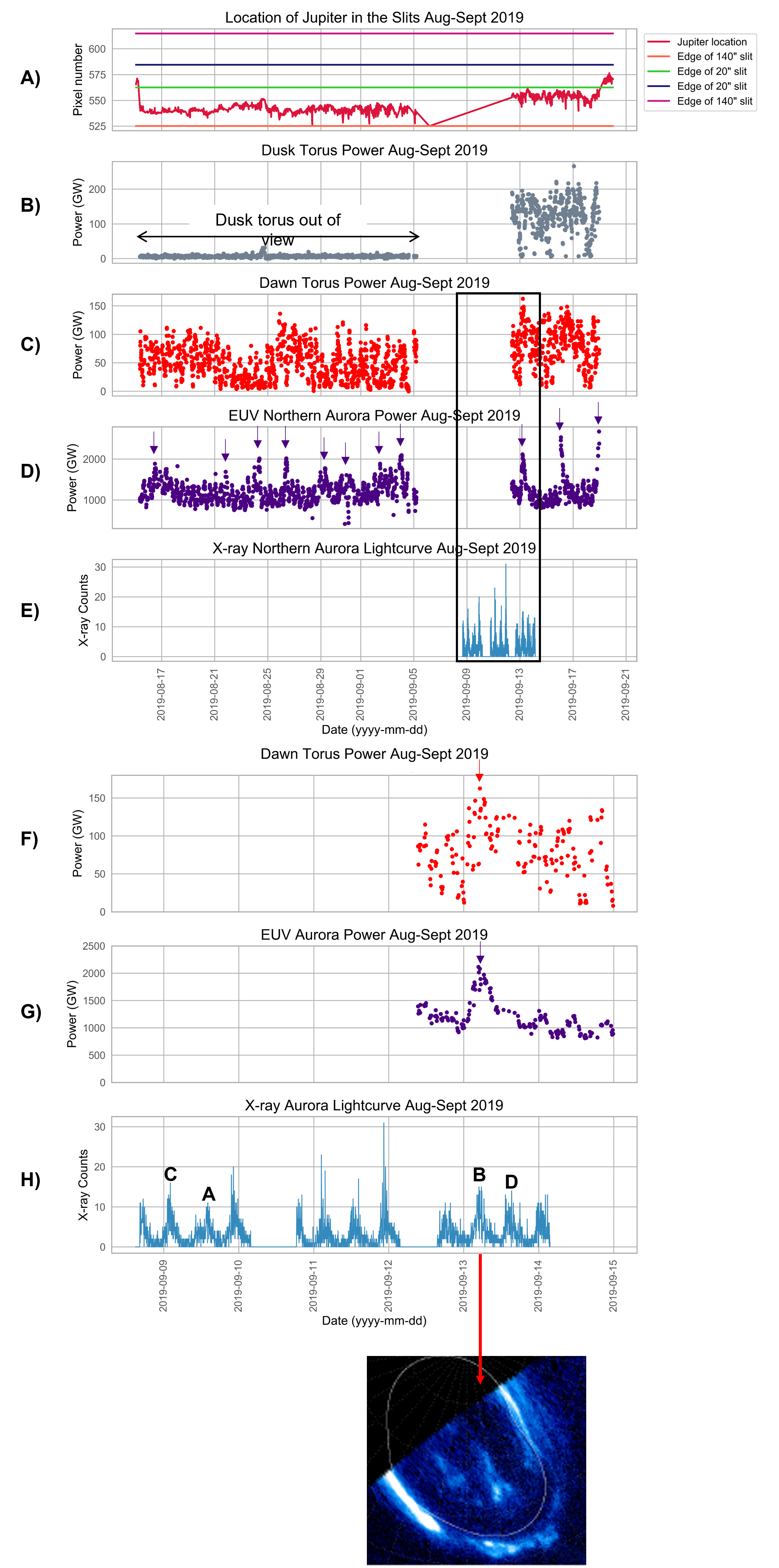}
    \caption{Panel A: Jupiter's location in the two slits; Panel B: Hisaki duskside torus power. Note that the duskside torus was out of view before 6 September and so no data was taken; Panel C: Hisaki dawnside torus power; Panel D: Hisaki northern EUV aurora power. The arrows show impulsive and transient brightenings; Panel E: XMM-Newton lightcurve of the northern X-ray aurora. No Hisaki data were taken between 7-11 September; Panels F-H: Zoom of black box from Panels C-E. The impulsive brightenings on 13 September in the dawnside torus and EUV aurora (red and purple arrows, respectively) matched in time with the appearances of the dawn storm and injection event in the FUV aurora. HST-observed Dawn Storm and Injection Events A and B, and Cases C and D are marked on the XMM-Newton lightcurve.}
    \label{fig:Hisaki}
\end{figure}

\subsection{IPT and Auroral Brightness}
The Hisaki satellite provides long term monitoring of Jupiter’s northern aurora and the IPT in EUV wavelengths and it allows us to see whether Io was loading mass into the jovian system. It is currently unclear how long it takes for the aurora to respond to a volcanic eruption from Io, however, previous studies (e.g. \cite{Kimura2018, Yoshikawa2017}) find that it is in the order of a few tens of days. We extracted the powers of the dawn and dusk sides of the IPT and the northern EUV aurora at 650-770 Å and 900-1480 Å (19-16 eV and 13-8 eV), respectively, from 15 August 2019 (24 days before XMM-Newton's observation) until 19 September (see panels B, C and D on Fig.~\ref{fig:Hisaki}). This is to ensure that we captured all of the volcanic activity that could have an effect on the aurora during XMM-Newton's observation period. Hisaki unfortunately suffered attitude problems which led to the loss of data between 7-11 September which means that we only have simultaneous EUV-X-ray data for the last 2 days of XMM-Newton's observation. Wavelengths between 650-770 Å are sensitive to the hot electron fraction in the central torus with energies of 10s-100s eV. Therefore, studying these wavelengths will give us a good indicator if there were any hot plasma injections.

Jupiter was located in Hisaki's 140 arcsec slit for the majority of this time interval (panel A on Fig.~\ref{fig:Hisaki}). This led to an overestimation on the auroral power before 12 September because of contamination from Jupiter's low latitude disc region. On the other hand, the dawn side torus brightness before 12 September is underestimated because the outer region of the dawn side torus was ``bitten out'' by the 20 arcsec slit. The dusk side torus was also outside of the slit and out of view to Hisaki. This prevents us from accurately determining the exact amount of mass that was loaded into the system and we also cannot compare the auroral and torus powers before and after 12 September. We can, however, look for relative time variabilities, such as impulsive and transient brightenings, in the powers.

%The torus brightness is a result of collisions between the ions and hot or cold electrons. The torus brightness could be increased by increasing its total mass content, or by keeping the total content constant but increasing the amount of hot electrons that it has. However, the torus brightness is likely to be proportional to the mass loading rate of the torus according to the analytical model described in \citeA{Kimura2018}. This would mean that the mass loading rate was higher after 12 September.  

The purple arrows in panel D of Fig.~\ref{fig:Hisaki} highlight impulsive brightening events in the northern EUV aurora. Their short-lived nature indicates that the active aurora was responding to internal drivers and their recurrence suggests that there was likely large mass loading from Io throughout the interval (\cite{Kimura2018}). Furthermore, the largest impulses occurred around and after the end of XMM-Newton's observation so it was likely that the magnetosphere was disturbed during the XMM-Newton observation.

Panels F, G and H in Fig.~\ref{fig:Hisaki} are zoom-ins of the black box that is over Panels C, D and E. The small impulsive brightening in the dawn torus on 13 September (shown by the red arrow), is a sign that a large scale injection penetrated the central torus between 6-9 R\textsubscript{J} and the large auroral impulse that occurred on the same day is a substorm-like event (see e.g. \cite{Bonfond2020}) that is usually seen during high mass loading periods. It is important to note that this substorm-like event is caused by magnetospheric configuration in Jupiter's magnetotail and is not associated with an Earth-like Dungey Cycle. The impulsive brightenings in the torus and the EUV aurora occur at the same time as Event B (bottom panel of Fig.~\ref{fig:Hisaki}) which means that those brightenings are associated with the dawn storms and plasma injection events that we see in the HST images.

\begin{figure}
    \includegraphics[width=\columnwidth]{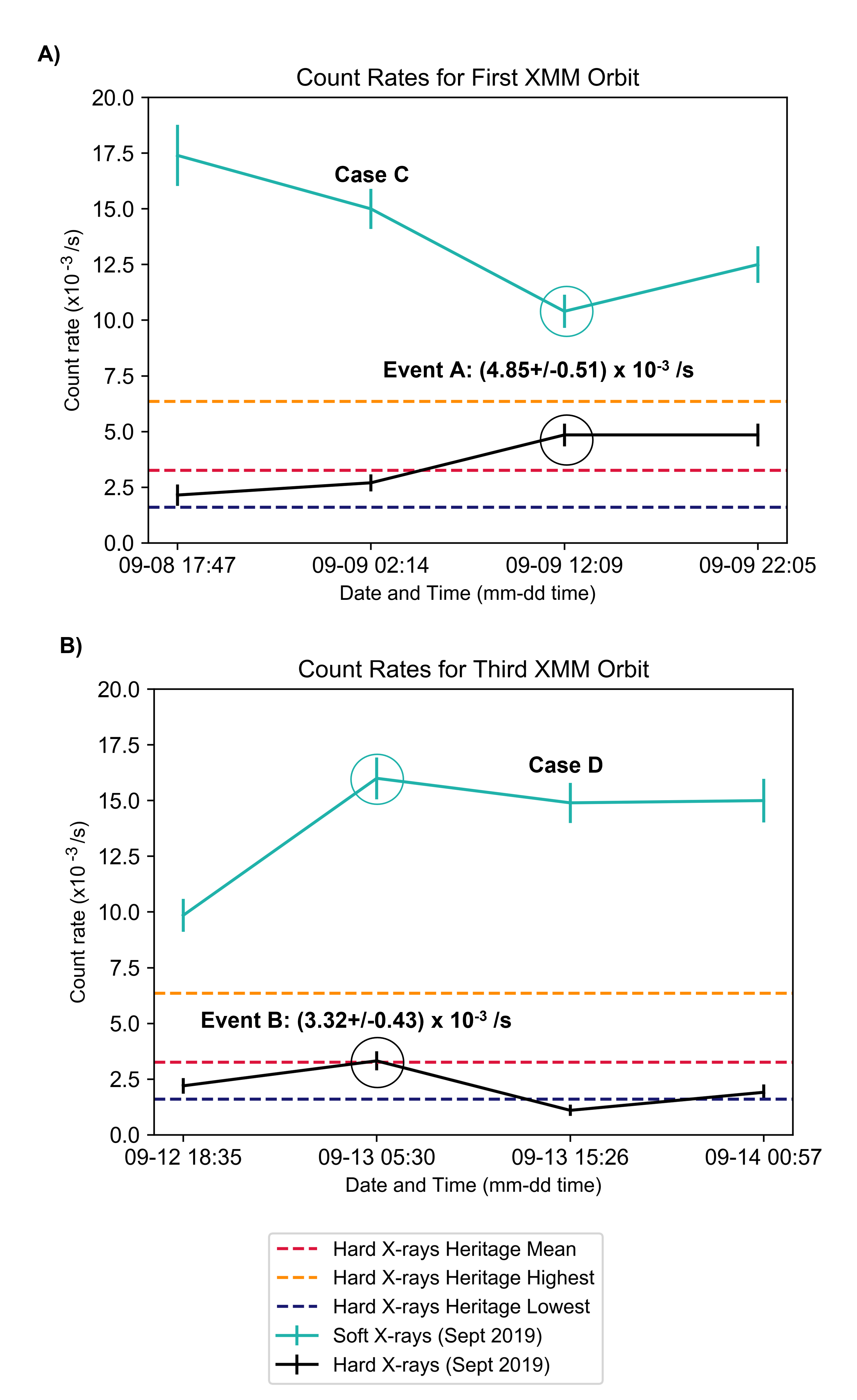}
    \caption{Panel A: Count rates during XMM-Newton's first orbit. The times shown on the x-axis are the midpoints of when the X-ray aurora was in view in UTC. Event A is highlighted by the circle and the time of Case C is shown. Panel B: Same as panel A but for XMM-Newton's third orbit and for Event B and Case D. Both plots show that the hard X-ray count rates (black line) during the September 2019 observations peaked when dawn storms and injection events are present. These values are also higher than the mean hard X-ray count rates (dashed red horizontal line) from observations taken between 2017-2019 after taking into account that Jupiter and XMM-Newton were separated by different distances during the different observations. The soft X-ray count rates (light green line) do not follow the same trend as the hard X-ray count rates during the first orbit, suggesting that an increase in electron precipitation does not automatically mean that there is also an increase in ion precipitation.}
    \label{fig:CountRates}
\end{figure}

We calculated the soft and hard X-ray count rates detected by the EPIC-pn instrument to determine how bright these emissions were every time the northern aurora was in view and the results are shown in Fig.~\ref{fig:CountRates} (and Table~\ref{table:SITable} in the Appendix). We have corrected the dependence of these values on the radial distance between Jupiter and XMM-Newton. The soft X-rays are produced by ion charge exchange, whilst the hard X-rays are due to electron bremsstrahlung. We also took the hard X-ray count rates from all eight observations taken between 2017-2019 and calculated the mean to be (3.26 $\pm$ 1.15)$\times$10$^{-3}$ s$^{-1}$. The lowest count rate during this period was (1.60 $\pm$ 0.18)$\times$10$^{-3}$ s$^{-1}$ while the highest was (6.35 $\pm$ 0.58)$\times$10$^{-3}$ s$^{-1}$. The peaks for emissions above 2 keV during XMM-Newton's first and third orbits in September 2019 were (4.85 $\pm$ 0.51)$\times$10$^{-3}$ s$^{-1}$ and (3.32 $\pm$ 0.43)$\times$10$^{-3}$ s$^{-1}$, respectively, and occurred at the same time as Events A and B (shown by the black circles). This result means that the FUV, EUV and hard X-ray auroral emissions, which are all caused by precipitating electrons, all brightened at the same time. The hard X-ray emissions remained high during the planetary rotation after Event A. HST did not take images of the northern FUV aurora for this rotation, but it did record a global brightening in the southern FUV aurora $\sim$4 hours after the northern aurora went out of view (Fig.~\ref{fig:HSTImages}). We cannot say when the global brightening started but it is possible that it did so when the northern aurora was in view and that this hemisphere was also experiencing the same brightening of the main oval which could explain why the hard X-ray emission was also bright.

Interestingly, the soft X-ray emissions, produced by precipitating ions, did not always follow the same trend as the hard X-rays as they brighten and dim at different times. The emissions below 2.0 keV also behave differently during Events A and B (light green circles). Event A had the lowest soft X-ray photon count during the first orbit, while Event B had the highest count rate for the same energy band during XMM-Newton's third orbit. It is intriguing that the soft and hard X-ray emissions for this orbit increase to their highest values during Event B, suggesting that there may have been an independent increase in the ion precipitation that happened at the same time as the electron precipitation. This indicates that the soft and hard X-ray aurorae can sometimes behave independently of each other and implies that the polar emissions (represented by the ions) also act independently of the dawn storm and injection events in the middle to inner magnetosphere.

\subsection{X-ray Spectral Fits} \label{XRaySpectra}
We extracted EPIC-pn spectra from Jupiter's northern auroral region during times when the aurora was in view. The aurora was visible 4 times for each XMM-Newton orbit and for $\sim$6 hours for most of those occasions (including Events A and B). The only exceptions were for the first and last times that the aurora was in view where they were visible for $\sim$3 and $\sim$5 hours, respectively. The data was binned so that each channel has at least 10 counts in order to balance the highest possible spectral resolution with robust spectral fitting statistics. We utilised the X-ray spectral fitting tool XSPEC (v. 12.10.1f released on 20 January 2020) and used the Atomic Charge Exchange (ACX) code (\cite{Smith}; http://www.atomdb.org/) and powerlaw models to find the best fit for each spectrum. ACX is used to model the charge exchange process leading to the auroral ionic emissions, whether the source of the ions is from the solar wind or from Io's volcanoes. ACX allows the user to input the ion abundances in an astrophysical plasma and the code outputs the charge exchange emission lines for that plasma at a given thermal energy, kT. This thermal energy dictates the charge states of the precipitating ions. The charge state distribution of the ions will change after each charge exchange process because the ions are able to keep charge exchanging until they are neutral. The ACX model assumes that the atmosphere that the precipitating ions collides with is cold and neutral and its H:He ratio can be assigned by the user. We set this ratio at 0.1 to represent Jupiter's atmosphere. However, the model does not account for hydrocarbons that are found in Jupiter's atmosphere, the viewing angle and associated optical thickness. For a complete review of the limitations of this approach see \cite{Dunn2020a}. The powerlaw model captures the tail in the spectrum above 2 keV and any unresolvable charge exchange lines at lower energies. 

We recreated the fast solar wind, slow solar wind and iogenic models in ACX described in \cite{Wibisono} and found that the iogenic plasma population of sulfur and oxygen ions, with the addition of one or two powerlaw continua gave the best fits for all of the northern X-ray auroral spectra. Table~\ref{table:FirstFits} and Table~\ref{table:ThirdFits} list all of the final parameter values for the best fits of each spectrum. The ACX temperature remained relatively stable during the observation period, with the exception of the slight increase on 09 Sep 02:14. In-situ measurements from different parts of Jupiter's magnetosphere by Voyagers 1 and 2, Ulysses, Galileo and Juno returned oxygen-to-sulfur (O-to-S) ratio values of 0.2-20.0 (\cite{Haggerty2019,Radioti}), while the physical chemistry model calculates it to be 1.02 (\cite{Delamere}). All of the O-to-S ratios in Table~\ref{table:FirstFits} and Table~\ref{table:ThirdFits} fall within the range found in the literature. %Both sets of spectra had comparable ACX normalization values but it is interesting that Event B has one of the highest ACX normalizations during the entire observation period. This means that the contribution from charge exchange was at one of its highest during this planetary rotation.

\begin{figure}
    \includegraphics[width=\columnwidth,height=0.8\textheight]{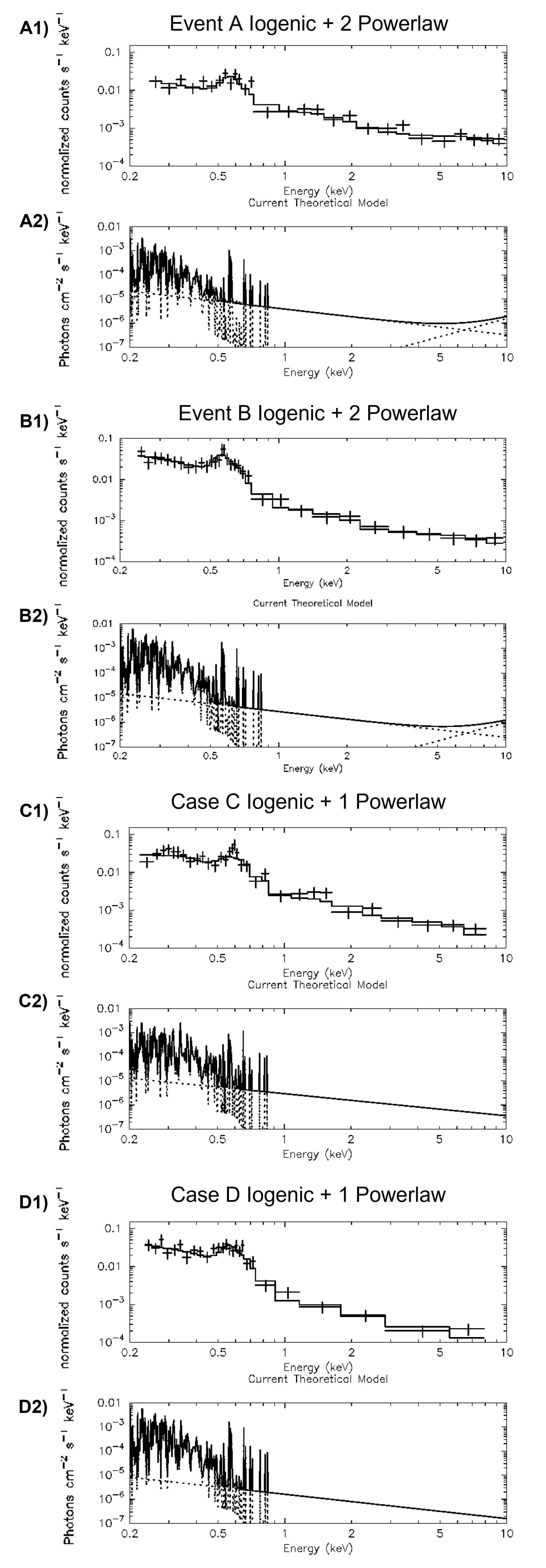}
    \caption{Crosses in panels A1, B1, C1 and D1 show datapoints of spectra extracted from Events A and B, and Cases C and D, respectively and the histograms show the best fits. Spectra A and B are for the planetary rotations with the dawn storms and injection events and have more hard X-ray (E \textgreater2 keV) emissions and enhanced tails than spectra C and D. Panels A2, B2, C2 and D2 show the theoretical models used to fit each spectrum with the solid lines displaying the dominant model and the dashed line the recessive model at a given energy.}
    \label{fig:Spectra}
\end{figure}

The powerlaw photon index, $\gamma$, determines the gradient of a spectrum as F(E)=E\textsuperscript{-$\gamma$}. A higher positive photon index gives a steeper slope than a lower one, while a negative photon index reverses the direction of the slope. Spectra with steep slopes imply that there are fewer energetic electrons precipitating and fewer hard X-ray photons are emitted. This gradient is most easily seen above 1 keV for jovian X-ray spectra as there are fewer charge exchange emission lines at these energies. Events A and B have more elevated tails as they have bright hard X-ray emissions (Fig.~\ref{fig:Spectra}). Both spectra have steep slopes below 5 keV but their bremsstrahlung tails are flatter at higher energies when compared with spectra without dawn storms and injection events. The first, steeper powerlaw may have been needed to present the blended emission lines at energies below 2 keV, while the second powerlaw model with the negative photon index was needed to fit the flat tails. This could also suggest that there was a second population of energetic electrons precipitating at these times. We include the spectral fits of Events A and B with the iogenic model and 1 powerlaw continuum in Figure Fig.~\ref{fig:A2} of the Appendix.

Table~\ref{table:FirstFits} and Table~\ref{table:ThirdFits} also list the fluxes and luminosity of the X-ray northern aurora taken during the two XMM-Newton orbits. The fluxes are obtained by integrating the area under their corresponding spectra (Fig.~\ref{fig:Spectra} and S2) as observed by XMM-Newton and the luminosity values are calculated by multiplying the fluxes with 4$\pi$r\textsuperscript{2} where r is the distance between Jupiter and XMM-Newton at the time. The results for both of the soft and hard X-rays in this study have a greater variation than what was quoted in \cite{Wibisono} as the luminosity from the aurora in June 2017 ranged between 0.32-0.37 GW for the soft X-rays and 0.13-0.15 GW for the hard X-rays. We also find that the X-ray aurora from September 2019 were brighter at both energy bands than those acquired 2 years previously for almost every planetary rotation. This is particularly true for Events A and B, and for the aurora witnessed one planetary rotation after Event A. All of these results agree with the findings from Fig.~\ref{fig:CountRates}.

\begin{table*}
	\centering
	\caption{Best fit spectral parameters for XMM-Newton's first orbit. The date and time (in UTC) shows the midpoints of when the northern X-ray aurora was in view}
	\label{table:FirstFits}
	\begin{tabular}{lcccc} % four columns, alignment for each
    \hline
    & \textbf{08 Sep 17:47} & \textbf{09 Sep 02:14 (Case C)} & \textbf{09 Sep 12:09 (Event A)} & \textbf{09 Sep 22:05} \\ [0.5ex] 
    \hline
    \textbf{Best fit model} & Iogenic & Iogenic & Iogenic & Iogenic\\
    \textbf{Reduced $\chi^{2}$} & 1.84 & 1.10 & 1.10 & 1.17 \\
    \textbf{Degrees of freedom} & 12 & 26 & 20 & 26 \\
    \textbf{ACX temperature (keV)} & 0.18 $\pm$ 0.01 & 0.24 $\pm$ 0.01 & 0.18 $\pm$ 0.01 & 0.19 $\pm$ 0.01 \\
    \textbf{ACX normalization ($\times10^{-6}$ photons cm\textsuperscript{-2} s\textsuperscript{-1} keV\textsuperscript{-1})} & 2.93 $\pm$ 0.44 & 1.37 $\pm$ 0.17 & 1.52 $\pm$ 0.27 & 1.22 $\pm$ 0.16\\
    \textbf{O-to-S ratio} & 0.80 $\pm$ 0.29 & 0.59 $\pm$ 0.17 & 0.83 $\pm$ 0.33 & 1.17 $\pm$ 0.40 \\
    \textbf{Powerlaw Photon Index 1} & 0.75 $^{+0.24}_{-0.33}$ & 0.92 $^{+0.18}_{-0.14}$ & 1.07 $^{+0.20}_{-0.15}$ & 0.42 $^{+0.13}_{-0.10}$ \\
    \textbf{Powerlaw normalization 1 ($\times10^{-6}$ photons cm\textsuperscript{-2} s\textsuperscript{-1} keV\textsuperscript{-1}))} & 2.23 $\pm$ 0.60 & 2.97 $\pm$ 0.51 & 3.82 $\pm$ 0.64 & 2.29 $\pm$ 0.34 \\
    \textbf{Powerlaw Photon Index 2} & - & - & -2.50 $^{+0.24}_{-0.20}$ & - \\
    \textbf{Powerlaw normalization 2 ($\times10^{-9}$ photons cm\textsuperscript{-2} s\textsuperscript{-1} keV\textsuperscript{-1}))} & - & - & 4.93 $\pm$ 2.10 & - \\
    \textbf{Flux for 0.2 - 2.0 keV ($\times10^{-5}$) photons cm\textsuperscript{-2} s\textsuperscript{-1})} & 15.89 & 10.30 & 8.30  & 8.25 \\
    \textbf{Flux for 2.0 - 10.0 keV ($\times10^{-5}$) photons cm\textsuperscript{-2} s\textsuperscript{-1})} & 0.53 & 0.54 & 1.01 & 0.91 \\
    \textbf{Luminosity for 0.2 - 2.0 keV (GW)} & 0.65 & 0.46 & 0.37 & 0.38 \\
    \textbf{Luminosity for 2.0 - 10.0 keV (GW)} & 0.34 & 0.33 & 0.75 & 0.61 \\
    \hline
 \end{tabular}
\end{table*}

\begin{table*}
	\centering
	\caption{Best fit spectral parameters for XMM-Newton's third orbit. The date and time (in UTC) shows the midpoints of when the northern X-ray aurora was in view.}
	\label{table:ThirdFits}
	\begin{tabular}{lcccc} % four columns, alignment for each
		\hline
		& \textbf{12 Sep 18:35} & \textbf{13 Sep 05:30 (Event B)} & \textbf{13 Sep 15:26 (Case D)} & \textbf{14 Sep 00:57} \\
		\hline
		\textbf{Best fit model} & Iogenic & Iogenic & Iogenic & Iogenic\\
		\textbf{Reduced $\chi^{2}$} & 0.78 & 0.52 & 0.99 & 0.92 \\
		\textbf{Degrees of freedom} & 16 & 25 & 21 & 19 \\
		\textbf{ACX temperature (keV)} & 0.21 $\pm$ 0.01 & 0.19 $\pm$ 0.01 & 0.19 $\pm$ 0.01 & 0.21 $\pm$ 0.01 \\
		\textbf{ACX normalization ($\times10^{-6}$ photons cm\textsuperscript{-2} s\textsuperscript{-1} keV\textsuperscript{-1})} & 1.20 $\pm$ 0.17 & 2.72 $\pm$ 0.31 & 2.64 $\pm$ 0.30 & 1.61 $\pm$ 0.19\\
		\textbf{O-to-S ratio} & 0.69 $\pm$ 0.24 & 0.64 $\pm$ 0.17 & 0.61 $\pm$ 0.16 & 0.93 $\pm$ 0.27 \\
		\textbf{Powerlaw Photon Index 1} & 0.41 $^{+0.19}_{-0.14}$ & 1.04 $^{+0.23}_{-0.17}$ & 1.01 $^{+0.29}_{-0.21}$ & 0.49 $^{+0.23}_{-0.17}$ \\
		\textbf{Powerlaw Photon Index 1} & 0.41 $^{+0.19}_{-0.14}$ & 1.04 $^{+0.23}_{-0.17}$ & 1.01 $^{+0.29}_{-0.21}$ & 0.49 $^{+0.23}_{-0.17}$ \\
        \textbf{Powerlaw normalization 1 ($\times10^{-6}$ photons cm\textsuperscript{-2} s\textsuperscript{-1} keV\textsuperscript{-1})} & 1.13 $\pm$ 0.24 & 2.76 $\pm$ 0.54 & 1.62 $\pm$ 0.39 & 1.14 $\pm$ 0.27\\
        \textbf{Powerlaw Photon Index 2} & - & -2.50 $^{+0.36}_{-0.23}$ & - & - \\
        \textbf{Powerlaw normalization 2 ($\times10^{-9}$ photons cm\textsuperscript{-2} s\textsuperscript{-1} keV\textsuperscript{-1})} & - & 3.14 $\pm$ 1.70 & - & - \\
        \textbf{Flux for 0.2 - 2.0 keV ($\times10^{-5}$) photons cm\textsuperscript{-2} s\textsuperscript{-1})} & 7.95 & 15.68 & 14.69 & 10.99 \\
        \textbf{Flux for 2.0 - 10.0 keV ($\times10^{-5}$) photons cm\textsuperscript{-2} s\textsuperscript{-1})} & 0.45 & 0.70 & 0.26 & 0.40 \\
        \textbf{Luminosity for 0.2 - 2.0 keV (GW)} & 0.34 & 0.65 & 0.59 & 0.48 \\
        \textbf{Luminosity for 2.0 - 10.0 keV (GW)} & 0.31 & 0.53 & 0.16 & 0.27 \\
        \hline
	\end{tabular}
\end{table*}

Fig.~\ref{fig:Spectra} presents the spectra for the planetary rotations that contain Events A and B, as well as two others that do not contain dawn storms or injection events for comparison (Cases C and D in Table~\ref{table:FirstFits} and Table~\ref{table:ThirdFits}). Spectra for the remaining planetary rotations can be found in Fig.~\ref{fig:A3} of the Appendix. The crosses in the upper panels show the data points. The lower panels show the model used to get the best fit and this model is convolved with the instrument response to produce the histogram in the upper panels. Jovian auroral spectra are dominated by a broad peak at 0.5-0.7 keV which is due to precipitating OVII and OVIII ions charge exchanging with native neutrals. The morphology of the spectra in Fig.~\ref{fig:Spectra} is similar to each other below 2 keV but there are distinct differences at higher energies. Spectra A and B have high levels of emissions above 2 keV giving them long and raised flat hard X-ray tails. Spectra C and D (and all of the X-ray spectra which had no coincident dawn storms and injection events) have fewer hard X-ray photon emissions and these photons are not detected at as high energies as those in Events A and B. This also explains why spectra for Events A and B have flatter tails at higher energies. Spectral studies of dawn storms in the FUV waveband also revealed that these features are associated with high energy electrons precipitating into Jupiter's atmosphere and a large amount of absorption due to CH\textsubscript{4} (\cite{Gustin2006}).

\subsection{X-ray Timing Analysis}
Jupiter's ionic soft X-ray aurorae have been widely reported to occasionally pulse with regular periods of tens of minutes (e.g. \cite{Gladstone,Jackman,Dunn2016}). It is thought that these pulsations are driven by ultra low frequency compressional mode waves found in the dawn to pre-midnight sectors of Jupiter's outer magnetosphere that then trigger electron and ion cyclotron (EMIC) waves in the plasma sheet and along the magnetic field lines. Particles located along the magnetic field lines interact with the EMIC waves and are pitch angle scattered into the planet's atmosphere to produce the bright auroral flares (\cite{Yao&Dunn}). The XMM-Newton lightcurves of emissions between 0.2-10.0 keV were rebinned to have 30s time-bins in order to increase the temporal resolution of the fast Fourier transform (FFT) analysis. We followed the method in \cite{Wibisono} to see whether the X-ray aurora had any quasi-periodic pulsations during Events A and B. The FFT power spectral density (PSD) plots for Events A and B are shown in Fig.~\ref{fig:FFTs} while the wavelet PSD plots are in the Appendix (Fig.~\ref{fig:A4}). Neither plots show any statistically significant quasi-periodic pulsations, therefore, it seems that tail reconnection events and their dipolarizations that are associated with dawn storms and injections, do not cause the ion pulsations and may even inhibit the regular periodicity of compressional mode waves in the outer magnetosphere. The only interval that showed strong pulsations during XMM-Newton's first and third orbits occurred on 9 September 20:00-22:00 UTC (the rotation after Event A) which had periods of $\sim$20 and $\sim$30 minutes. The FFT PSD plots for the rest of the times that the X-ray aurora was visible are shown in Fig.~\ref{fig:A5}. The temporal results also show, at least during these observations, that there is no clear connection between processes happening in the outer magnetosphere that are responsible for the pulsed behaviour, with those that are in the middle magnetosphere which produce the dawn storms and injection events.

\begin{figure}
    \includegraphics[width=\columnwidth]{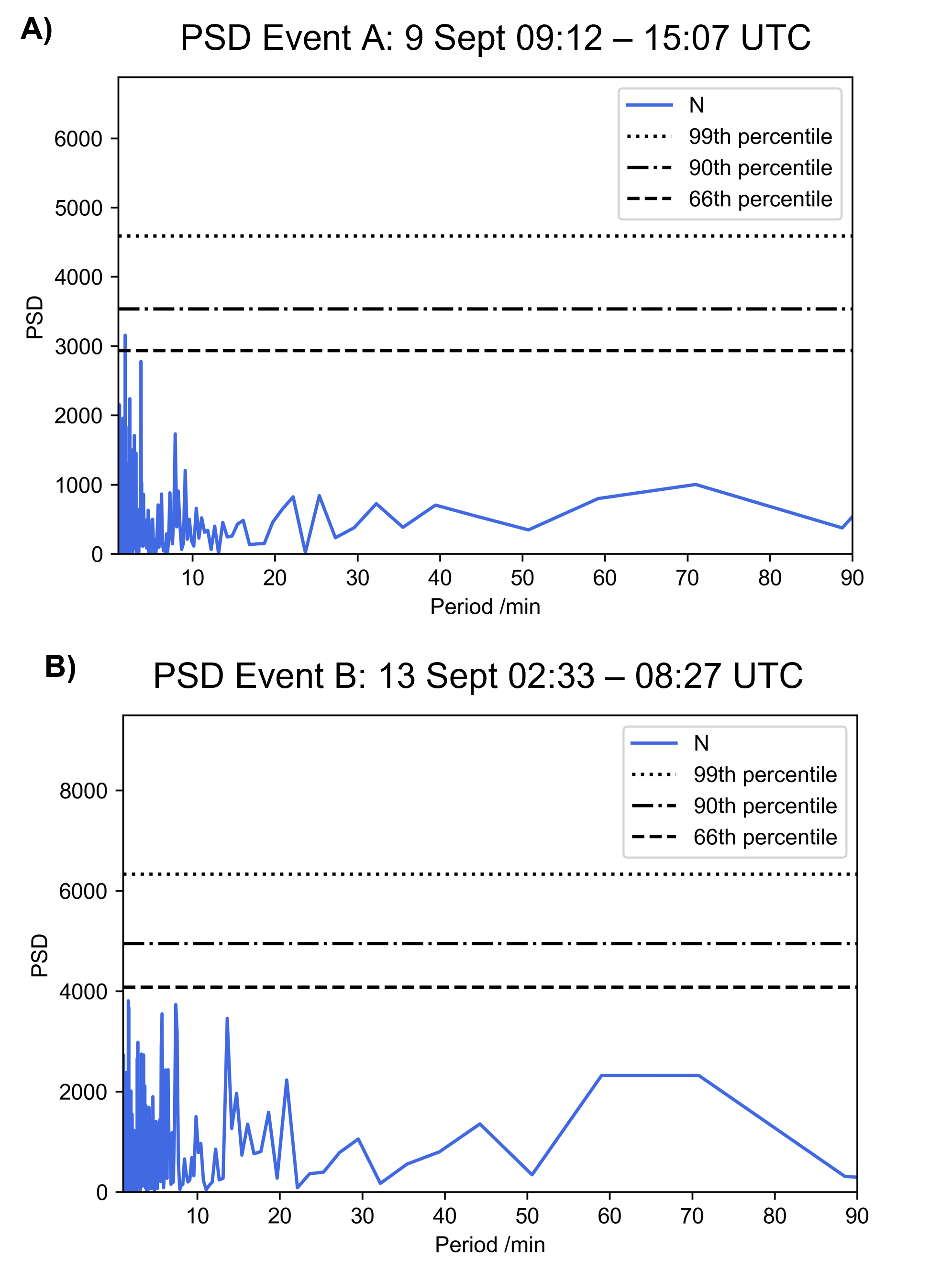}
    \caption{Fast Fourier transform power spectral density plots for the planetary rotations containing Events A (Panel A) and B (Panel B). The black horizontal dashed, dashed-dot and dotted lines show the 66th, 90th and 99th percentiles respectively. No statistically significant quasi-periodic pulsations were found during these two intervals.}
    \label{fig:FFTs}
\end{figure}

\section{Discussion}
Results from a multiwavelength campaign by XMM-Newton, HST and Hisaki in September 2019 are presented in this study. HST images taken concurrently with XMM-Newton observations show that the FUV aurora was affected by two sets of dawn storms and injection events and we study their consequences for the X-ray aurora during this observation period. These phenomena appeared and disappeared within one Jupiter rotation. The northern EUV aurora had impulsive and frequent brightenings in the weeks prior to, during, and after the observation period.  Hisaki observed a clear short-lived brightening in the dawnside IPT that coincided with Event B which is evidence that a large amount of plasma was injected into the central torus in the inner magnetosphere and caused the dawn storm and injection event on the same day. Spectral analysis of the X-ray northern aurora shows that the precipitating ions were predominantly from an iogenic source. All of these features indicate that the aurora was responding to internal processes rather than those driven by the solar wind.

There was no solar wind monitoring upstream of Jupiter at the time of the observations and the planet was far from opposition with the Earth (the Earth-Jupiter angle relative to the Sun was \textgreater60\textdegree) which meant that solar wind propagation models, such as the one described in \cite{Tao} (see Fig.~\ref{fig:A6}), would be expected to give inaccurate results in the arrival time of solar wind shocks at Jupiter. Nevertheless, according to the model, two small solar wind compressions arrived at Jupiter in the week before XMM-Newton's observations and the solar wind exerted a dynamic pressure of $\sim$8.5 nPa on Jupiter's magnetosphere on 8 September - one day before Event A and 5 days before Event B. Although those shocks would have had some impact on the aurora, the internal drivers are more likely to be what caused the dawn storms and injection events to appear and produce the results seen by Hisaki and XMM-Newton. This is because internal drivers cause features that only last for one Jupiter rotation or so. External drivers, such as solar wind compressions, on the other hand, can leave their mark on the aurora over several Jupiter rotations.

Event A was missed by Hisaki due to the limitations of the instrument but Event B was captured by all three observatories. The northern FUV, EUV and hard X-ray aurorae increased in brightness at this time which suggests that the magnetic reconnection and subsequent dipolarization of the field lines led to more electrons to accelerate into Jupiter's atmosphere. It appears that ion precipitation was not affected as the soft X-ray count rates did not always follow the same trend as shown by the hard X-ray emissions. FFT analysis of the XMM-Newton lightcurves show that the ionic X-ray aurora did not pulse quasiperiodically during Events A and B. This was also the case for five of the six auroral viewing windows that did not have the dawn storms and injections seen in the FUV aurora. Therefore, it appears that, at least for some cases, the hard and soft X-rays are driven by different processes at different parts of Jupiter's magnetosphere that work independently of each other, in agreement with \cite{GBRSpec}.

The results from the X-ray spectral fits show that the precipitating ions have an iogenic origin. Io has hundreds of active volcanoes on its surface. Every second they release $\sim$1000 kg of predominantly neutral SO$_{2}$ into the moon's vicinity. Roughly a third to a half of the ejected SO$_{2}$ dissociate and are ionised through collisions and photoionisation. The newly formed S and O ions are picked up by Jupiter's magnetic field and are accelerated from 17 km s\textsuperscript{-1} to 74 km s\textsuperscript{-1} to bring them to corotation with the planet (\cite{Bagenal2017}). Centrifugal forces push the ionised volcanic material outwards and flatten it into a plasma sheet with the most dense plasma forming the IPT. \cite{Cravens2003} explains the current system needed to accelerate magnetospheric ions and electrons into Jupiter's atmosphere to produce the aurorae. Furthermore, observational results from e.g. \cite{Cravens,Dunn2016,Wibisono} also led to conclude that the ions responsible for the X-ray aurora are predominantly from Io's volcanoes. Theoretical and modelling studies agree with those findings and show that precipitating high energy state S and O ions charge exchanging with native neutrals in Jupiter's atmosphere can produce Jupiter's auroral soft X-rays (\cite{Cravens,Hui2009,Hui2010,Ozak,Ozak2013,Houston2020}). All of this supports our conclusion that the X-ray emissions observed during the dawn storms and injections of Events A and B have their ultimate origin in Io's volcanic activity. The X-ray spectra also show hints that during dawn storms and injections, a second population of energetic electrons is what causes the aurora to release high energy (\textgreater5 keV) X-ray photons which gives these spectra elevated and flat bremsstrahlung tails.

\section{Conclusions}

Studying Jupiter's northern FUV, EUV and X-ray aurorae simultaneously during the presence of dawn storms and injection events revealed a number of new findings and supported some ideas that are already in the literature:
\begin{enumerate}
  \item The aurora in all three wavebands increased in brightness when dawn storms and injections appeared which means that there must have been an increase in energetic electrons precipitating into Jupiter's atmosphere, or that the precipitating electrons were more energetic.
  \item The low and high energy X-ray emissions behave independently to each other suggesting that there is an independency between processes occurring in the outer magnetosphere (diagnosed through the soft X-rays) with those happening in the middle and inner magnetosphere (reflecting hard X-ray electron processes). Therefore, there is also an independency between ion and electron precipitation. This is shown by the results from the X-ray count rates and the timing analysis.
  \item X-ray spectra of the aurora with dawn storms and injections have long and flat bremsstrahlung tails which are best fit by one powerlaw model with a positive photon index and a second with a negative index.
  \item X-ray spectral analysis finds that the soft end of the spectra is best fit with a model that consists of iogenic ions suggesting that the source of the precipitating ions is predominantly originally from Io's volcanoes. 
\end{enumerate}

\section*{Acknowledgements}

A.D.W and R.P.H are supported by the Science and Technology Facilities Council (STFC) (Project nos. 2062546 and 2062537, respectively). A.J.C., G.B.R. and W.R.D. acknowledge support from STFC consolidated grant ST/S000240/1 to University College London (UCL). D.G. acknowledges the financial support from the Belgian Federal Science Policy Office (BELSPO) via the PRODEX Programme of ESA. Z.H.Y. acknowledges the Key Research Program of the Institute of Geology \& Geophysics, CAS, Grant No. IGGCAS- 201904. T.K. was supported by a Grant-in-Aid for Scientific Research KAKENHI (20H01956, 20KK0074, 19H01948, 19H05184) from the Japan Society for the Promotion of Science (JSPS). H.K. was supported by Grant‐in‐Aid for JSPS Research Fellow and KAKENHI 
(19H01948). We thank Chihiro Tao for her 1 dimensional magnetohydrodynamic model that propagates the solar wind from Earth to Jupiter.

%%%%%%%%%%%%%%%%%%%%%%%%%%%%%%%%%%%%%%%%%%%%%%%%%%
\section*{Data Availability}
 
The UV auroral images are based on observations with the NASA/ESA Hubble Space Telescope (program HST GO-15638), obtained at the Space Telescope Science Institute (STScI), which is operated by AURA for NASA. All data are publicly available at STScI https://mast.stsci.edu/portal/Mashup/Clients/Mast/Portal.html. Ephemeris to see when the X-ray aurora was in view were created from the NASA JPL HORIZONS web-interface https://ssd.jpl.nasa.gov/horizons.cgi. Hisaki data is archived in the Data Archives and Transmission (DARTS) JAXA https://darts.isas.jaxa.jp/stp/hisaki/. The raw and calibrated XMM-Newton data can be downloaded from the XMM-Newton Science Archive http://nxsa.esac.esa.int/nxsa-web/\#home. We used the XMM-Newton Science Analysis Software (SAS) (https://www.cosmos.esa.int/web/xmm-newton/download-and-install-sas), XSPEC (https://heasarc.nasa.gov/docs/xanadu/xspec/) and Atomic Charge Exchange (ACX) (http://atomdb.org/) to extract and analyse the auroral spectra.



\bibliographystyle{mnras}
\bibliography{example} % if your bibtex file is called example.bib

%\begin{thebibliography}{99}
%\end{thebibliography}

%%%%%%%%%%%%%%%%%%%%%%%%%%%%%%%%%%%%%%%%%%%%%%%%%%

%%%%%%%%%%%%%%%%% APPENDICES %%%%%%%%%%%%%%%%%%%%%

\appendix

\section{EPIC-pn images of Jupiter}
\begin{figure}
    \includegraphics[width=\columnwidth]{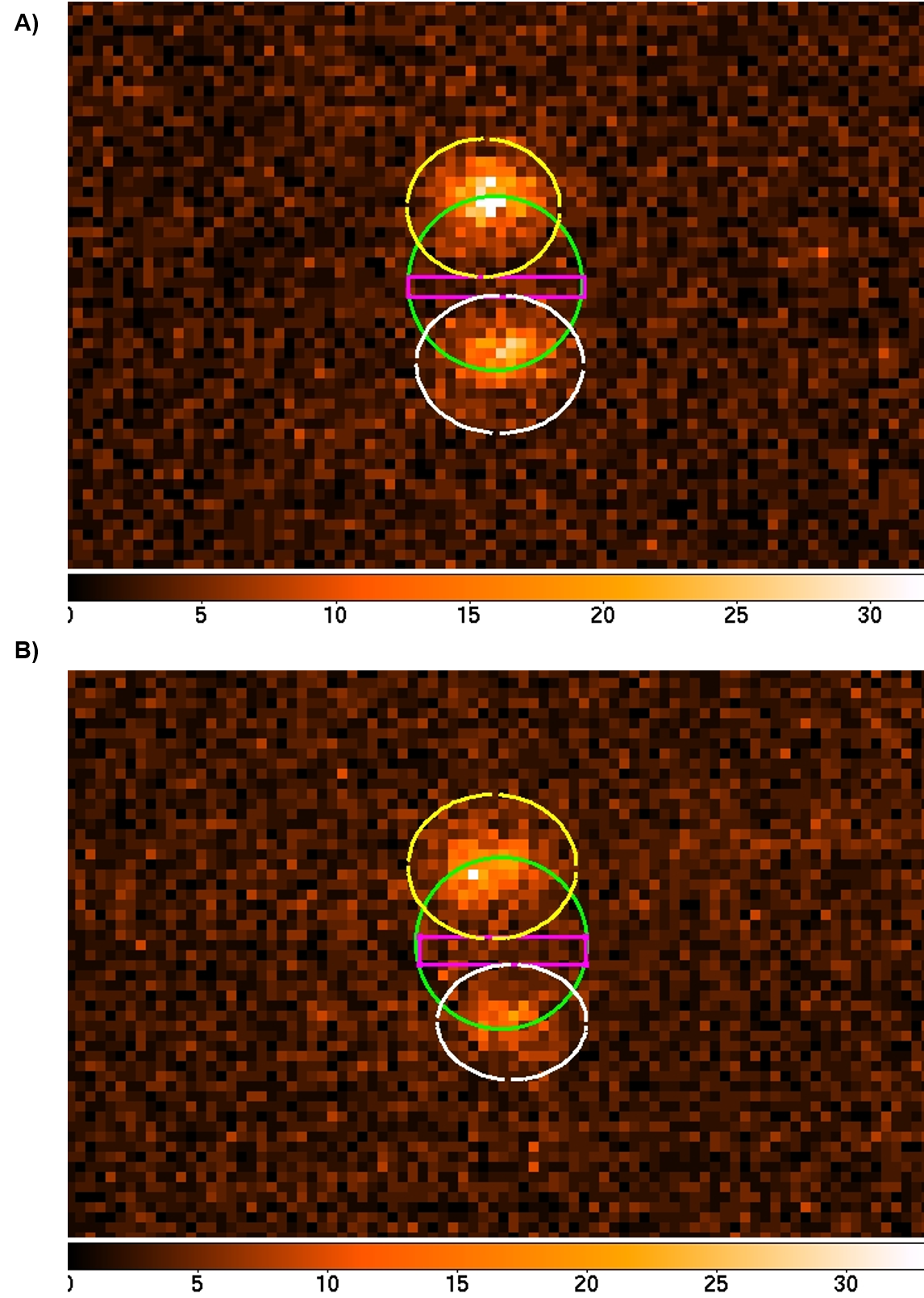}
    \caption{XMM-Newton EPIC-pn images of Jupiter from the spacecraft’s first (panel A) and third (panel B) orbits. The northern aurora region is highlighted by the yellow oval, the southern aurora by the white oval and the equatorial region by the pink rectangle. Data were extracted from these regions to produce the lightcurves and spectra presented in the main text. Jupiter’s disc is shown by the green circle. The angular diameter of Jupiter was 38.0 arcsec for the first orbit and 37.6 arcsec for the third orbit. The aurorae extend beyond Jupiter’s disc in these images because of the blurring introduced by XMM-Newton’s relatively low spatial resolution. The colourbar shows the number of X-ray photon counts in each pixel.}
    \label{fig:A1}
\end{figure}

Background X-ray sources appear stationary in the sky during XMM-Newton's observations. Therefore, these sources remain fixed on the instruments' detectors. Jupiter's motion is obvious due to it being much closer, therefore, the planet will move across the detectors and appear as a streak in images. The X-ray photons can be re-registered into a Jupiter-centered co-ordinate system so that Jupiter now appears fixed in the image while the static background sources streak across the image. Fig.~\ref{fig:A1} shows the Jupiter-centered image of the planet's X-ray aurorae taken by the EPIC-pn instrument during XMM-Newton's first and third orbits. Regions were drawn over the auroral and equatorial regions to determine where data for the spectral and timing analyses are to be extracted. 

\section{X-ray photon count rates}

We found from Fig.~\ref{fig:CountRates} that the soft and hard X-ray emissions brightened and dimmed at different times, and that the hard X-ray aurora brightened simultaneously with the FUV and EUV aurorae. Tab.~\ref{table:SITable} lists the count rates of X-ray photons in both energy ranges.

\begin{table*}
	\centering
	\caption{Count rates of X-ray photons emitted by Jupiter for XMM-Newton’s first and third orbits. The date and times show the midpoints of when the northern X-ray aurora was in view. We have accounted for the different distances between Jupiter and XMM-Newton between each observation. The graphs for this table are found in Fig.~\ref{fig:CountRates} in the main text.}
	\label{table:SITable}
	\begin{tabular}{lcc}
		\hline
		\textbf{Date and Time (UTC)} & \textbf{Count rates of soft X-rays ($\times10^{-3}$ s\textsuperscript{-1})} & \textbf{Count rates of hard X-rays ($\times10^{-3}$ s\textsuperscript{-1})}\\
		\hline
		08 Sep 17:47 & 17.40 $\pm$ 1.37 & 2.15 $\pm$ 0.48 \\
		09 Sep 02:14 & 15.00 $\pm$ 0.90 & 2.70 $\pm$ 0.38 \\
	    09 Sep 12:09 (Event A) & 10.40 $\pm$ 0.75 & 4.85 $\pm$ 0.51 \\
	    09 Sep 22:05 & 12.50 $\pm$ 0.82 & 4.85 $\pm$ 0.51 \\
	    12 Sep 18:35 & 9.85 $\pm$ 0.74 & 2.20 $\pm$ 0.35 \\
	    13 Sep 05:30 (Event B) & 16.00 $\pm$ 0.94 & 3.32 $\pm$ 0.43 \\
	    13 Sep 15:26 & 14.90 $\pm$ 0.90 & 1.10 $\pm$ 0.25 \\
	    14 Sep 00:57 & 15.00 $\pm$ 0.98 & 1.91 $\pm$ 0.35 \\
		\hline
	\end{tabular}
\end{table*}

\section{Best fit spectra}

\begin{figure}
    \includegraphics[width=\columnwidth,height=0.8\textheight]{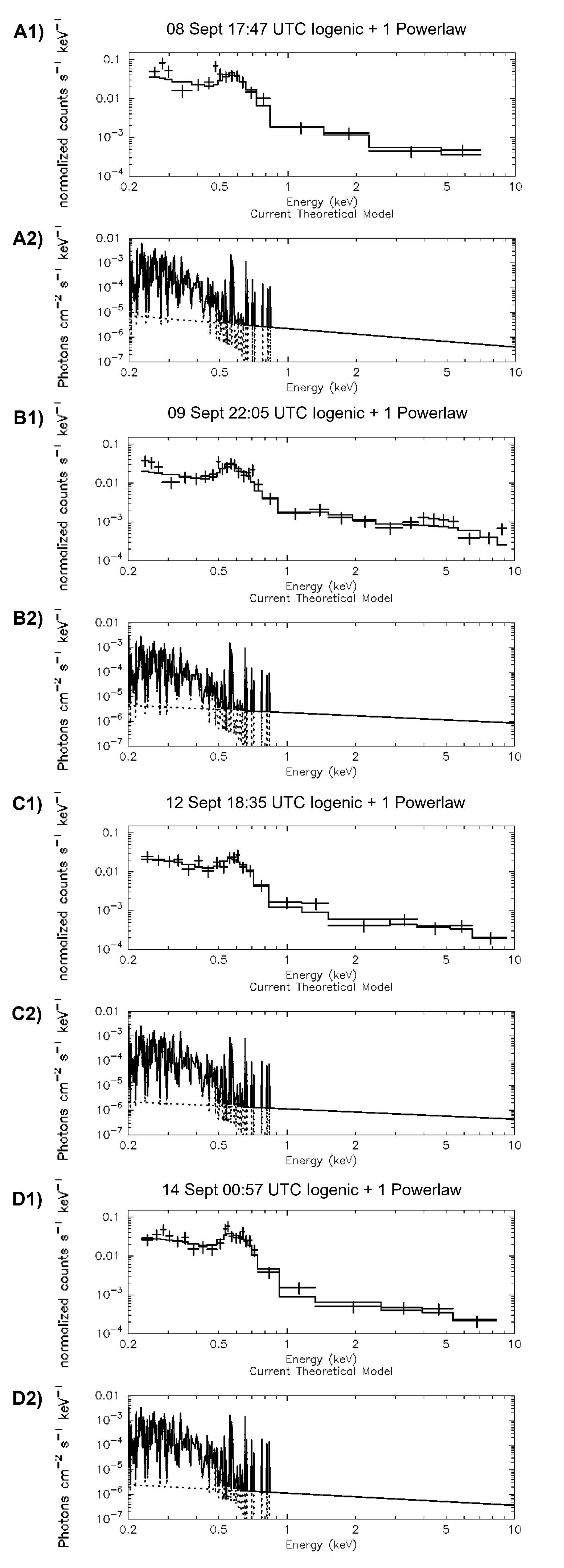}
    \caption{The best spectral fits for the northern X-ray aurora. The fits are represented by the histograms and the data points by the crosses in the upper panels (A1, B1, C1, D1). The theoretical models are shown in the lower panels (A2, B2, C2, D2) and the solid and dashed lines display the dominant and recessive model respectively at any given energy. The date and time above each spectrum are the midpoints of when the northern X-ray aurora was in view. The best fit parameters are found in Tab.~\ref{table:FirstFits} and Tab.~\ref{table:ThirdFits} in the main text.}
    \label{fig:A2}
\end{figure}

Spectra were extracted each time that the northern X-ray aurora was in view and best fitted using the Atomic Charge Exchange (ACX) code (\cite{Smith}) in XSPEC. Fig.~\ref{fig:A2} shows the best spectral fits for the planetary rotations that are not shown in Fig.~\ref{fig:Spectra}.

\section{Spectra for Events A and B with one powerlaw}

\begin{figure}
    \includegraphics[width=\columnwidth]{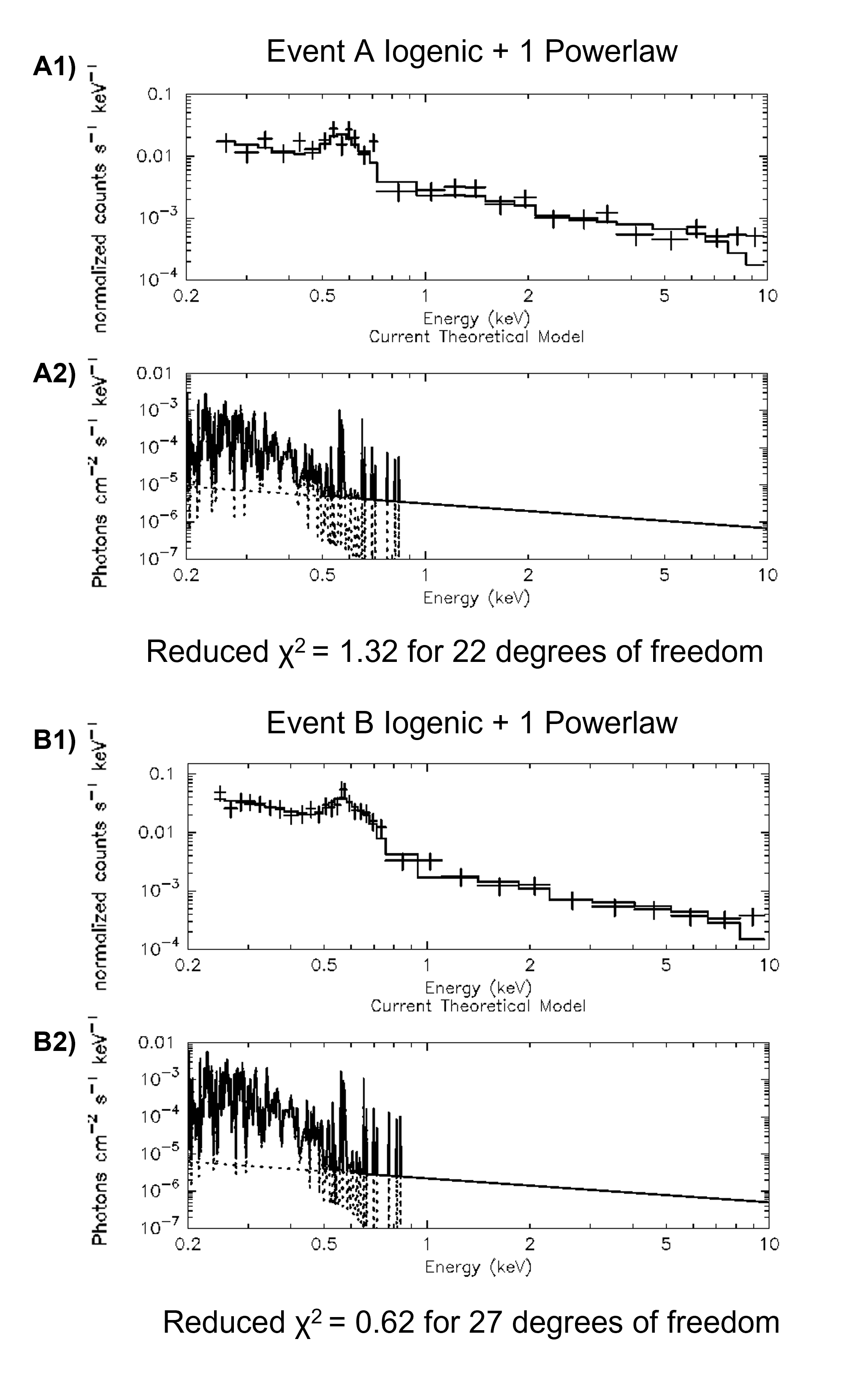}
    \caption{Spectral fits for Events A and B with the iogenic and 1 powerlaw models. These fits are not as good as what was achieved when the second powerlaw continuum was added which helped to fit the spectra at higher energies. The fits are represented by the histograms and the data points by the crosses in the upper panels (A1, B1). The theoretical models are shown in the lower panels (A2, B2) and the solid and dashed lines display the dominant and recessive model respectively at any given energy.}
    \label{fig:A3}
\end{figure}

Events A and B are occasions when the northern X-ray aurora had coincident dawn storms and injections in the FUV aurora. The X-ray spectra for these were best fitted with a model consisting of an iogenic ion population and 2 powerlaw continua to capture the enhanced bremsstrahlung tails and unresolvable charge exchange lines. Fig.~\ref{fig:A3} show that 1 powerlaw continuum gave worse fits for these spectra. 

\section{Wavelet power spectral density plots}

\begin{figure}
    \includegraphics[width=\columnwidth]{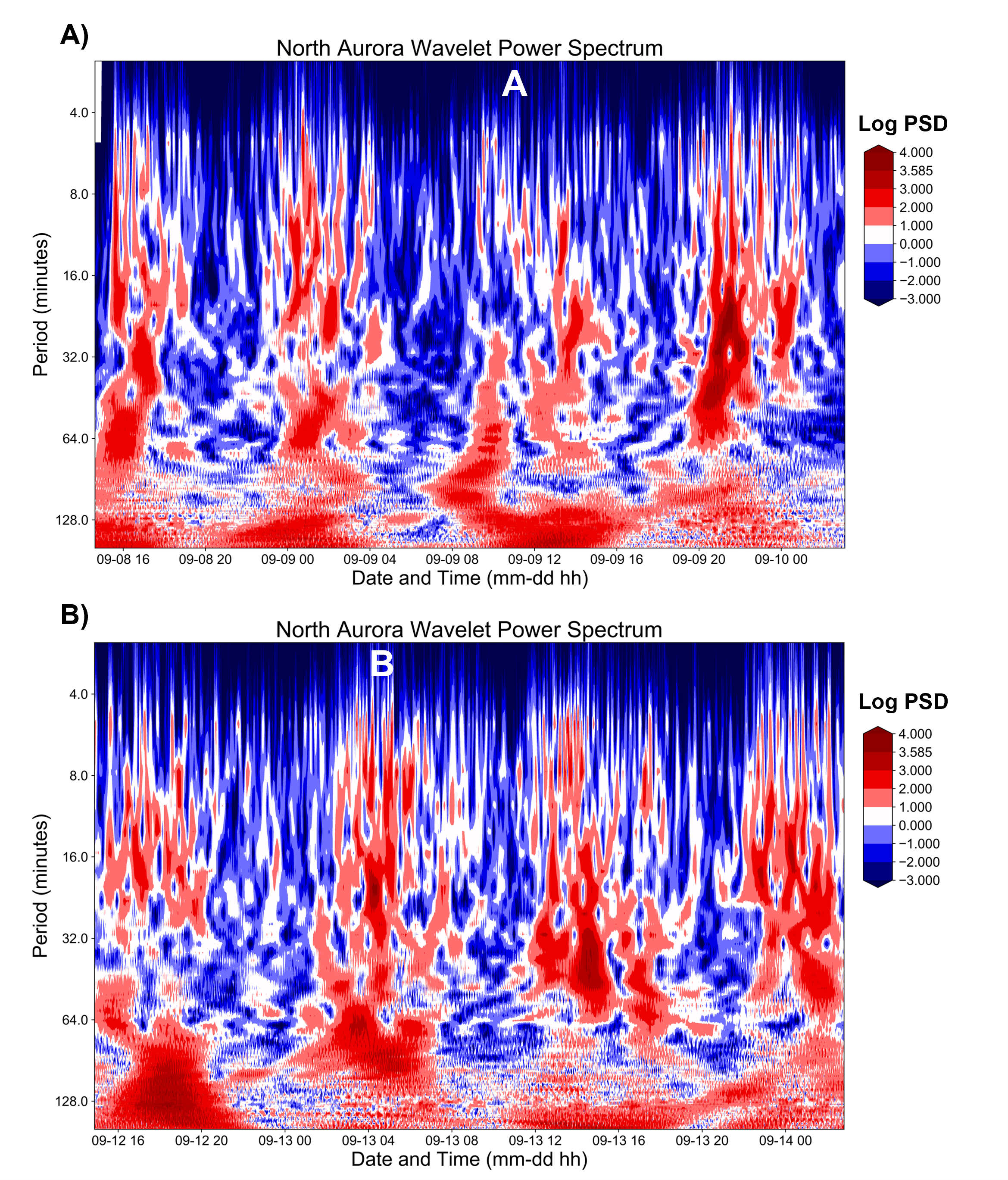}
    \caption{Power spectral density (PSD) plots for the northern aurora with 2-min time resolutions during XMM-Newton’s A) first orbit and B) third orbit using the Shannon wavelet. The colour bar shows PSD on a log scale from 2\textsuperscript{-3} to 2\textsuperscript{4}. Areas in dark red have strong quasi-periodic pulsations. The only time interval to show this was on 9 September 20:00-22:00 UTC. Events A and B are marked on. The X-ray aurora was visible during the times in red. These intervals, plus the dark red regions on 9 September 20:00–22:00 UTC and 13 September 14:00–16:00 UTC were further analysed using the fast Fourier transform method.}
    \label{fig:A4}
\end{figure}

We used a wavelet transform method as described in \cite{Wibisono} to determine time intervals when Jupiter's soft X-ray aurora had strong pulsations. This method cannot give the time and period of the pulsations accurately, but it does give a visual presentation of them. The power spectral density (PSD) plots in Fig.~\ref{fig:A4} give n us estimates of time intervals that need further analyses to determine the behaviour of the pulsations.

\section{Fast Fourier transform power spectral density plots}

\begin{figure}
    \includegraphics[width=\columnwidth]{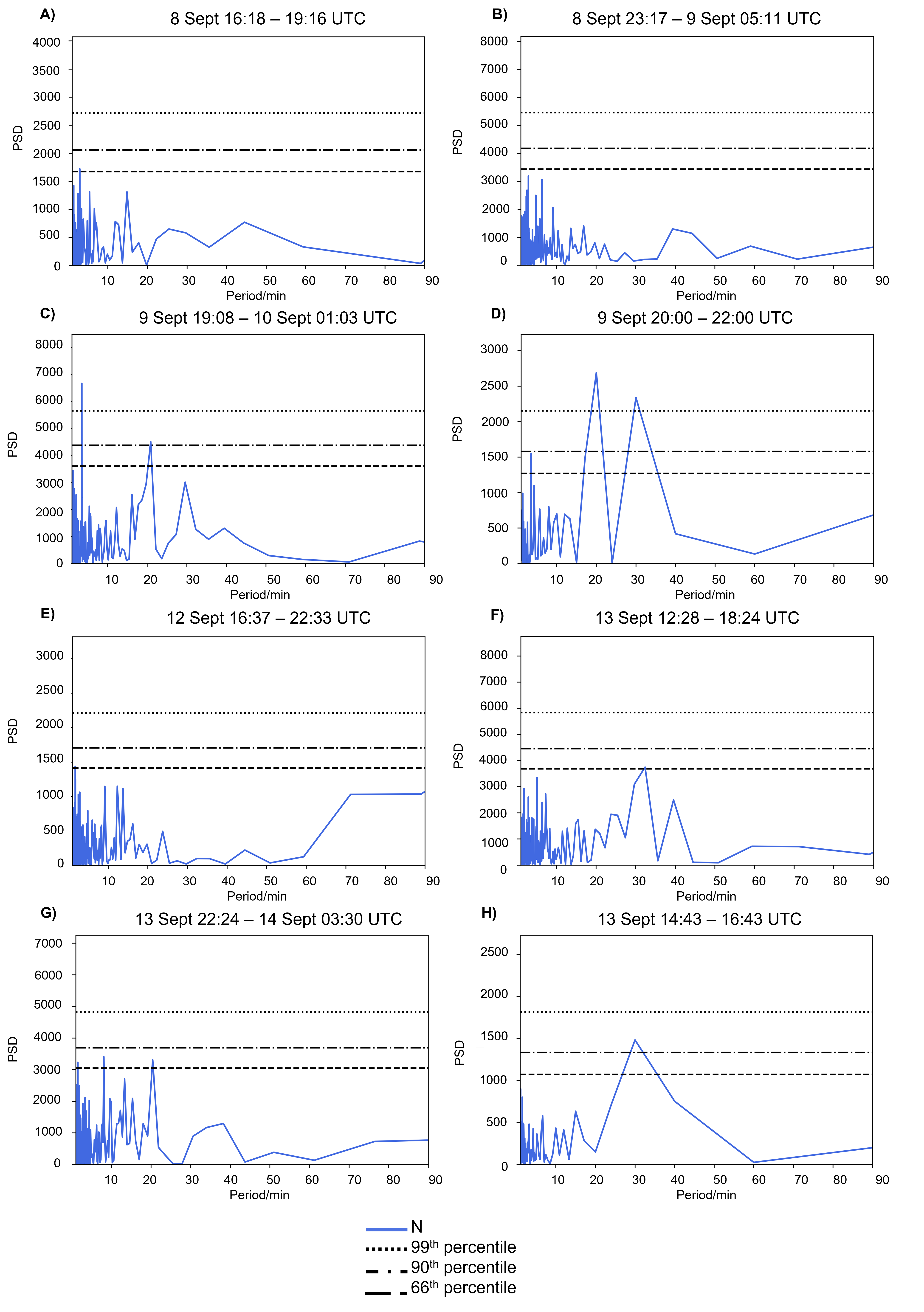}
    \caption{Fast Fourier transform PSD plots for each time that the northern X-ray aurora was in view and did not have dawn storms or injection events during XMM-Newton’s first orbit (A, B, C) and third orbit (E, F, G). PSD D and H are the intervals mentioned in Fig.~\ref{fig:A4}. Panel D shows the only PSD that had statistically significant pulsations over this observation period.}
    \label{fig:A5}
\end{figure}

A fast Fourier transform (FFT) was applied over the time intervals when the northern X-ray aurora was in view and those that were identified by the wavelet transform method. Unlike the wavelet transform, the FFT does not give simultaneous resolutions in time and period. Therefore, the periods of pulsations can be more accurately determined. The PSDs from the FFT analysis that were not included in the main text are shown in Fig.~\ref{fig:A5}.

\section{Solar wind propagation model}

\begin{figure}
    \includegraphics[width=\columnwidth]{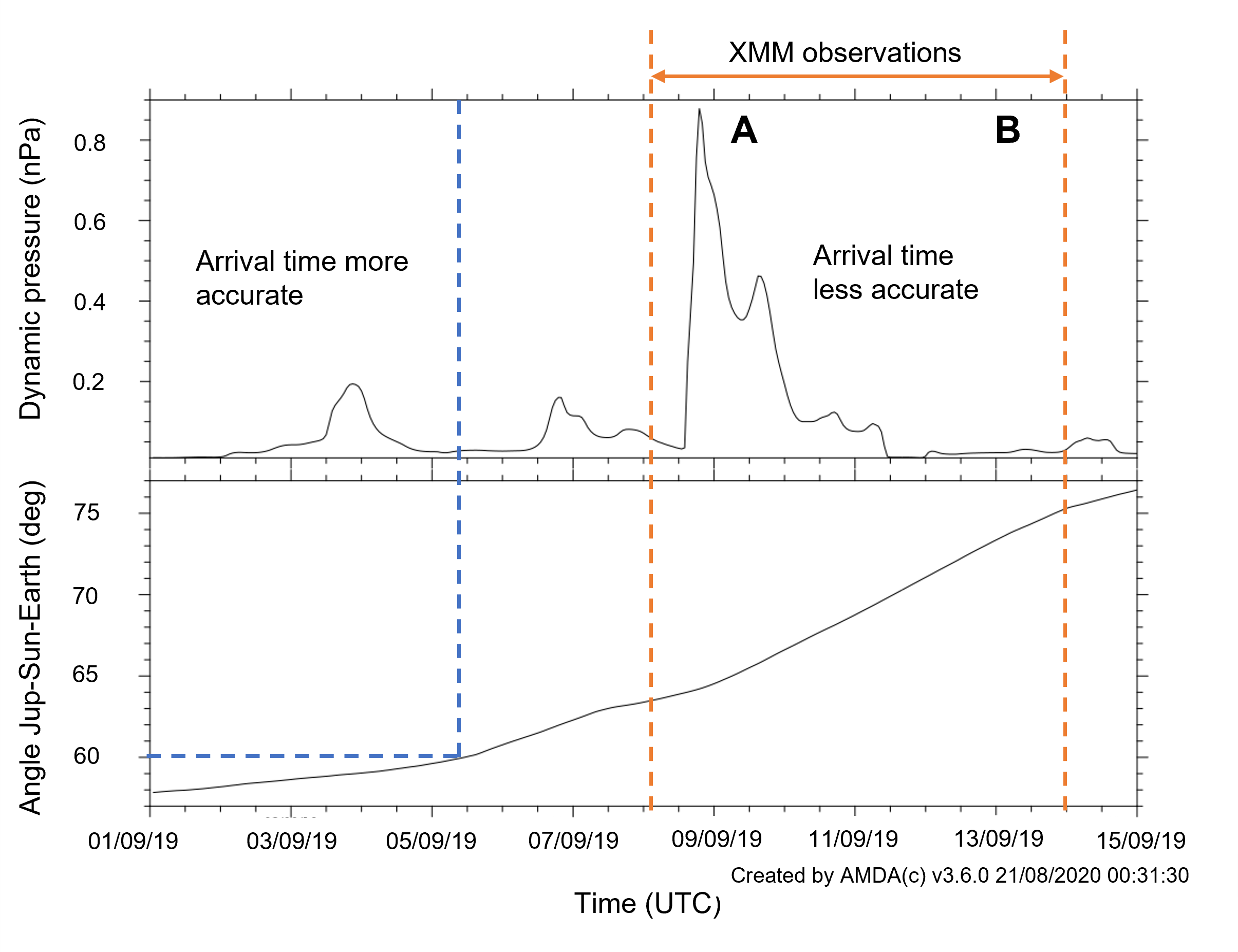}
    \caption{The propagation model of the solar wind dynamic pressure using the \protect\cite{Tao} one-dimensional magnetohydrodynamic model (top panel). It shows that a few small solar wind shocks arrived at Jupiter the week before XMM-Newton started observing the planet. A large solar wind shock with a dynamic pressure of 0.85 nPa hit Jupiter on 8 September. However, the angle between the Earth and Jupiter relative to the Sun (bottom panel) was larger than 60° at this time which meant that the error of this arrival time is at least $\pm$2 days. The arrival time for the first small shock was more accurate as this angle was below 60° (boundary is shown by the dashed blue line). The dashed orange lines mark when XMM-Newton started and stopped observing. Events A and B are shown in this figure. The model can be accessed from 
    http://amda.cdpp.eu/. }
    \label{fig:A6}
\end{figure}

Jupiter's X-ray aurora can be influenced by the solar wind. For example, the emissions brighten when a large solar wind dynamic pressure is exerted on the planet's magnetosphere. Studies also suggest that solar wind compressions of the magnetosphere may help to trigger the quasi-periodic pulsations of the soft X-ray aurora (\cite{Dunn2016,Wibisono}). Fig.~\ref{fig:A6} shows the solar wind parameters estimated by the \cite{Tao} model in the weeks before, and during this study's observation period.

%%%%%%%%%%%%%%%%%%%%%%%%%%%%%%%%%%%%%%%%%%%%%%%%%%

% Don't change these lines
\bsp	% typesetting comment
\label{lastpage}
\end{document}